\documentclass[preprint,aps,showpacs]{revtex4}

\usepackage{graphicx}

\begin{document}

\title{Comment on ``'t Hooft vertices, partial quenching, and rooted
  staggered QCD''}
\author{Michael Creutz}
\affiliation{
Physics Department, Brookhaven National Laboratory\\
Upton, NY 11973, USA
}

\begin{abstract}
{A recent criticism of the proof of the failure of the rooting
procedure with staggered fermions is shown to be incorrect.}
\end{abstract}

\maketitle

In a recent paper \cite{Bernard:2007eh} Bernard, Golterman, Shamir and
Sharp challenge the proof developed in
\cite{Creutz:2007yg,Creutz:2007rk,Creutz:2007yr} showing that
non-perturbative effects are incorrectly treated in the rooting
formalism popular for reducing the number of fermion species in the
staggered formalism for dynamical quarks.  Here I discuss how this
challenge is based on a misunderstanding of the chiral
behavior of staggered quarks.

The problem appears already in the introduction of
\cite{Bernard:2007eh} where the authors attempt to define something
they call the ``rooted continuum theory'' (RCT).  While technically
the paper makes the qualifying statement ``What is less certain ... is
that the rooted staggered theory on the lattice becomes this RCT as
$a\rightarrow 0$,'' the discussion misleadingly continues as if the
RCT can be defined in two inequivalent ways.  First they consider the
continuum limit of staggered fermions treated with the rooting trick.
But then towards the end of the introduction they say the RCT can be
obtained rigorously from rooting four copies of a chirally invariant
formulation, such as with the overlap operator \cite{overlap}.  For
the latter theory the correctness of rooting is a trivial mathematical
identity.

The point of the discussion in
Refs.~\cite{Creutz:2007yg,Creutz:2007rk,Creutz:2007yr} is that these
two approaches display qualitatively different non-perturbative
effects.  Only the latter form generates the correct one flavor
theory.  Confusing these theories is equivalent to assuming that
rooting is correct and misses the issues that invalidate the rooting
prescription when used with staggered quarks.  Throughout the rest of
the paper they make no distinction between these definitions, just
referring to the RCT as the physical one flavor theory.  If the RCT
operator is chosen from rooting four equivalent copies of a properly
defined chiral fermion theory, then the remaining discussion in
Ref.~\cite{Bernard:2007eh} is simply a verification of the trivial
reduction to the one flavor case.

The crucial issue is that after rooting the staggered propagator still
represents four independent fermion states.  A valid single fermion
propagator would have only a single pole.  While rooting does
correctly reweight perturbative loops, it fails when instantons are
present.  Then the 't Hooft vertex \cite{'tHooft:fv} directly couples
all fermion species, including any extra tastes.  Rooting four powers
of a true single fermion theory involves a propagator that has only
one physical pole.  Coupling four copies of this pole via instantons
is impossible due cancellations from Pauli statistics.  This
cancellation does not occur for the staggered tastes which remain as
independent states, leaving an incorrect form for the 't Hooft vertex.
The undesired effects occur at a typical instanton scale, which is set
by $\Lambda_{\rm QCD}$, and will survive the continuum limit.  The
problems appear whenever instanton physics is important, irrespective
of whether the quark masses vanish.

The introduction to Ref.~\cite{Bernard:2007eh} also propagates the
misconception that it is only taste breaking and mixing that can
cause problems.  The issue with rooting is not taste
breaking but the strong coupling between the tastes induced by these
non-perturbative effects.  The troublesome coupling takes the form of
a determinant that is in fact taste symmetric.  Since the coupling
involves all tastes, it appears only in processes roughly comparable
to four loops in the perturbative expansion.  Thus these effects may
not be large in flavor non-singlet processes.  But their existence
rules out the method as a first principle approach to physical
observables.

That taste breaking and the strong coupling between tastes are
independent issues has recently been explored in
Ref.~\cite{Adams:2008db}, where a two taste model for rooting is
constructed with taste mixing explicitly removed.  The arguments of
Refs.~\cite{Creutz:2007yg,Creutz:2007rk,Creutz:2007yr} still apply,
and rooting fails in this model also because the two tastes involved
are physically inequivalent.  The model is similar to Wilson fermions
but considers one taste with an effective the strong interaction CP
angle theta of $\pi/2$ and the second of $-\pi/2$, as discussed in
Ref.~\cite{Seiler:1981jf}.  This rotation allows a residual chiral
symmetry to survive at finite lattice spacing, but does not commute
with the rooting process.  Considered as a two flavor theory, these
phases cancel, but on rooting one is working with a mixture of two
inequivalent one-flavor theories.  Ref.~\cite{Adams:2008db} does not
commit on whether this make sense, but argues that if it does, the
theory will not have CP violation and therefore must be the one flavor
theory at vanishing theta.  However, from a fundamental point of view,
rooting the product of two inequivalent theories is not in general
expected to make sense.

The authors of Ref.~\cite{Bernard:2007eh} proceed to rewrite the
partition function for their theory in terms of a partially quenched
approach with three ghost fields.  Again they do not distinguish the
RCT Dirac operator used and assume they can use the same one for each
field, including the ghosts.  Actually, Ref.~\cite{Creutz:2007rk} also
raises the possibility of using ghost fields to reduce the flavor
content of unrooted staggered quarks, but emphasizes the important
proviso that the ghosts must be formulated with a chiral operator,
such as the overlap, to properly cancel the inequivalent tastes.

In summary, Ref.~\cite{Bernard:2007eh} confuses two different rooted
theories, one of which is correct and the other not.  The distinction
is a strong inter-taste coupling that survives in the continuum limit
and does not allow the effects of a single taste to be isolated.  The
successes of past simulations do suggest that these effects can be
small for some observables, but it is incorrect to claim that they go
away in the continuum limit.  Finally, I note that the argument that
staggered simulations are much faster than alternative approaches has
recently become moot \cite{DelDebbio:2006cn}.

\section*{Acknowledgements}
This manuscript has been authored under contract number
DE-AC02-98CH10886 with the U.S.~Department of Energy.  Accordingly,
the U.S. Government retains a non-exclusive, royalty-free license to
publish or reproduce the published form of this contribution, or allow
others to do so, for U.S.~Government purposes.

\end{document}